\documentclass[prl,twocolumn,superscriptaddress]{revtex4}

\usepackage{graphicx}
\usepackage{dcolumn}
\usepackage{bm}
\usepackage{natbib}
\usepackage{amsmath}
\usepackage{amssymb}
\usepackage{here}

\begin{document}
\preprint{preprint - not for distribution}

\title{InAs nanowire hot-electron Josephson transistor}

\author{Stefano Roddaro}
\author{Andrea Pescaglini}
\author{Daniele Ercolani}
\author{Lucia Sorba}
\author{Francesco Giazotto}
\affiliation{NEST, Istituto Nanoscienze-CNR and Scuola Normale Superiore, Piazza S. Silvestro 12, I-56127 Pisa, Italy}
\author{Fabio Beltram}
\affiliation{NEST, Istituto Nanoscienze-CNR and Scuola Normale Superiore, Piazza S. Silvestro 12, I-56127 Pisa, Italy}
\affiliation{IIT@NEST, Center for Nanotechnology Innovation, Piazza S. Silvestro 12, I-56127 Pisa, Italy}

\date{\today}

\begin{abstract}
At a superconductor (S)-normal metal (N) junction pairing correlations can ``leak-out'' into the N region. This {\em proximity effect}~\cite{deGennes,Andreev64} modifies the system transport properties and can lead to supercurrent flow in SNS junctions~\cite{ABS}. Recent experimental works showed the potential of semiconductor nanowires (NWs) as building blocks for nanometre-scale devices~\cite{Lieber03,Bjork02,Roddaro08,Linke09}, also in combination with superconducting elements~\cite{Xiang06,Vandam06,Doh05,Schapers09,kasper}. Here, we demonstrate an InAs NW Josephson transistor where supercurrent is controlled by hot-quasiparticle injection from normal-metal electrodes. Operational principle is based on the modification of NW electron-energy distribution~\cite{Wilhelm,Morpurgo98,Schaepers98,Baselmans99,Savin04,Giazotto04,Crosser06,Tirelli08} that can yield reduced dissipation and high-switching speed. We shall argue that exploitation of this principle with heterostructured semiconductor NWs opens the way to a host of out-of-equilibrium hybrid-nanodevice concepts \cite{Linke09,Giazotto06}.
\end{abstract}

\maketitle

\begin{figure}[th!]
\begin{center}
\includegraphics[width=0.48\textwidth]{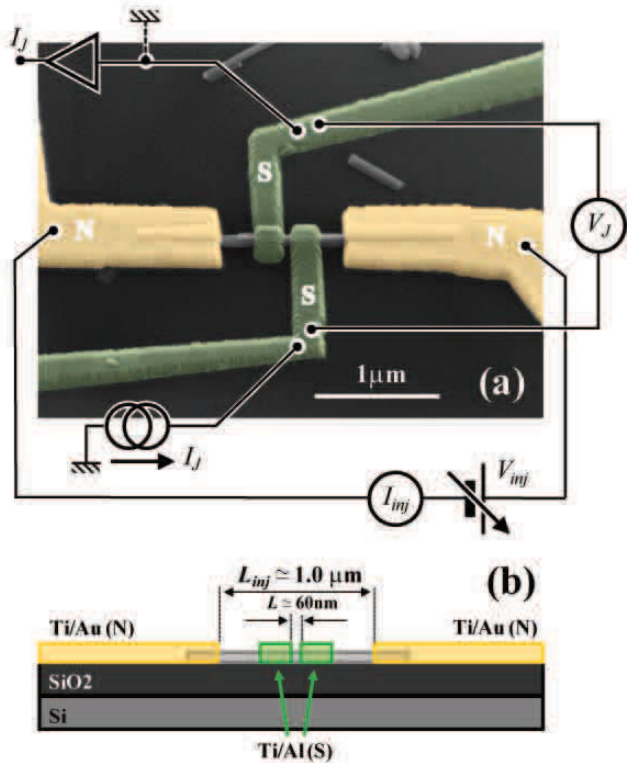}
\caption{{\bf InAs NW hot-electron Josephson transistor}. (a) Pseudo-color scanning electron micrograph of one of the measured devices: an InAs NW is coupled to two $250\,{\rm nm}$-wide Ti/Al superconducting electrodes (green) to form a $\simeq 60\,{\rm nm}$-long Josephson junction. The InAs NW is also contacted at its ends to two normal-metal Ti/Au leads (yellow) which are used to control the electron energy distribution in the central region of the junction through quasiparticle injection. As a consequence, the Josephson supercurrent $I_J$ is tuned upon voltage biasing the N control line with $V_{inj}$. The distance between the two normal-metal electrodes is $L_{inj}\simeq 1\,{\rm \mu m}$. The overlay schematics shows a sketch of the measurement set-up: the supercurrent is detected using a four-wire scheme, while the injection circuit consists of a floating network. (b) Schematic cross-section view of a typical device with some relevant geometrical details. The transistors were fabricated on top of an insulating Si/SiO$_2$ substrate by electron-beam lithography and metal deposition.}
\end{center}
\end{figure}
One practical implementation of the present InAs-NW hot-electron Josephson transistor concept is shown in Figure~1a. The SNS junction was fabricated by e-beam lithography starting from an n-doped InAs NW (the N region, for further details see Methods) and comprises two Ti/Al superconducting electrodes placed at a distance $L\simeq60\,{\rm nm}$ (S regions). Control electrodes were fabricated by depositing two additional normal-metal Ti/Au leads at the two ends of the NW. As schematically illustrated by the overlay of Fig.~1a, the N leads are used to drive a dissipative current through the NW thereby tuning Josephson coupling in the S-NW-S structure.

\begin{figure}[t!]
\begin{center}
\includegraphics[width=0.5\textwidth]{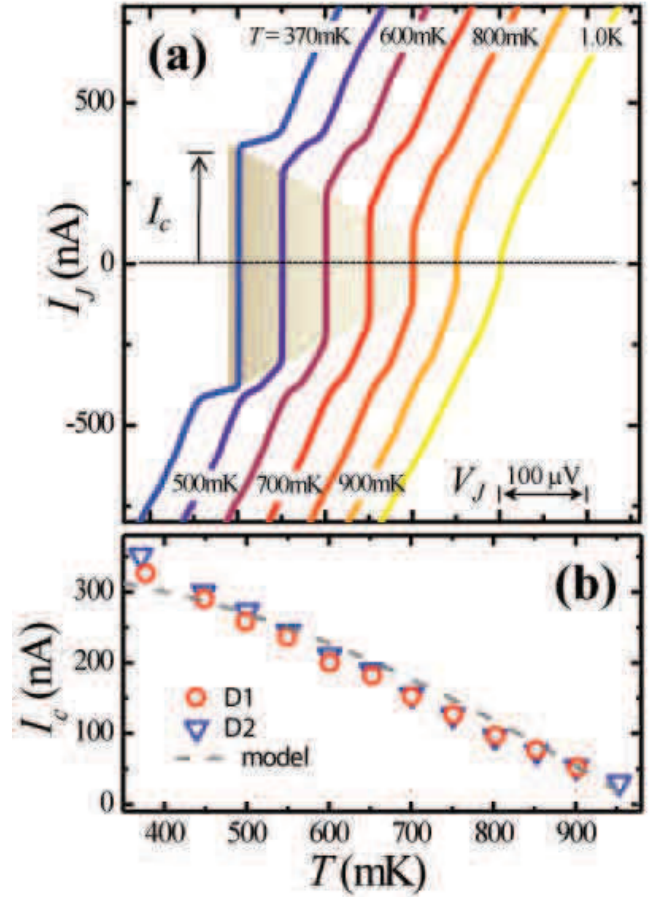}
\caption{{\bf Temperature dependence of the Josephson current}.
(a) The current-voltage characteristics $I_J$ vs $V_J$ of device D2 demonstrate the presence of dissipationless transport up to a temperature $T\simeq1\,{\rm K}$. $I_c$ is the critical current of the junction, and the hatched area provides a guide to the eye to emphasize the supercurrent suppression as a function of temperature. The curves are horizontally shifted for clarity.
(b) Evolution of the measured equilibrium critical current $I_c$ as a function of bath temperature $T$ for two representative devices (D1-red circles, D2-blue triangles). Dashed line is the theoretical prediction for an ideal short diffusive SNS junction. The absence of a clear saturation of $I_c(T)$ suggests that higher critical currents could be observed at lower temperature. The junction normal-state resistance is $R_N\simeq 220\,\Omega$ and $\simeq 210\,\Omega$ for device D1 and D2, respectively.
}
\end{center}
\end{figure}
Figure~2a shows a set of typical IV characteristics from one of the measured S-NW-S junctions at equilibrium (i.e., $V_{inj}=0$) and at different bath temperatures $T$. High critical currents up to $I_c \simeq 350\,{\rm nA}$ (corresponding to a supercurrent density $\sim 5.5\,{\rm kA/cm^2}$) are observed and Josephson coupling typically persists up to about $1\,{\rm K}$. Furthermore, from the junction differential-resistance spectra (see Supplementary Materials) we infer a superconducting order parameter $\Delta = 120\,{\rm \mu eV}$ and a junction normal-state resistance $R_N \simeq 210\,{\rm \Omega}$. The $I_cR_N$ product attains values as large as $\simeq 75\,{\rm \mu eV}$ and indicates the overall success of our junction-fabrication procedure~\cite{Heikkila}. Based on carrier-density and electron-mobility values (see Methods) we estimate a momentum-relaxation length $\ell\simeq20\,{\rm nm}$ and a diffusion coefficient $D=0.02\,{\rm m^2/s}$. Given the junctions geometrical length $L$ we can estimate the Thouless-energy value, i.e., the characteristic energy scale of the N region, $E_{T\!h}=\hbar D/L^2\simeq 4\,{\rm meV}$. These values indicate that our devices can be described within the frame of the diffusive short-junction limit ($L>\ell$ and $\Delta\ll E_{T\!h}$) \cite{Heikkila}. Figure~2b shows the evolution of $I_c$ versus $T$ at $V_{inj}=0$ for two representative devices D1 and D2. For comparison, the theoretical prediction for an ideal short diffusive SNS junction~\cite{Heikkila,ShortDiffusive} is also plotted assuming for both devices a supercurrent suppression of the order of $45\%$. Such a suppression can be expected and probably mainly stems from a residual barrier at the NW-superconductor contact~\cite{Heikkila,Courtois08,Garcia09}.

Figure~3 demonstrates operation of the NW hot-electron Josephson transistor: supercurrent can be fully suppressed by the application of a voltage bias $V_{inj}$ of the order of a few hundred $\,{\rm \mu V}$ across the N control line. The overall evolution of the IV characteristics of device D2 as a function of $V_{inj}$ at $370\,{\rm mK}$ is displayed in Fig.~3a. The corresponding monotonic decay and complete suppression of the critical current $I_c$ vs $V_{inj}$ at different bath temperatures is reported in Figs.~3b and 3c for devices D1 and D2, respectively. Our results show that an injection power of barely $\sim 100\,{\rm pW}$ is sufficient to completely quench the supercurrent.
\begin{figure}[t!]
\begin{center}
\includegraphics[width=0.5\textwidth]{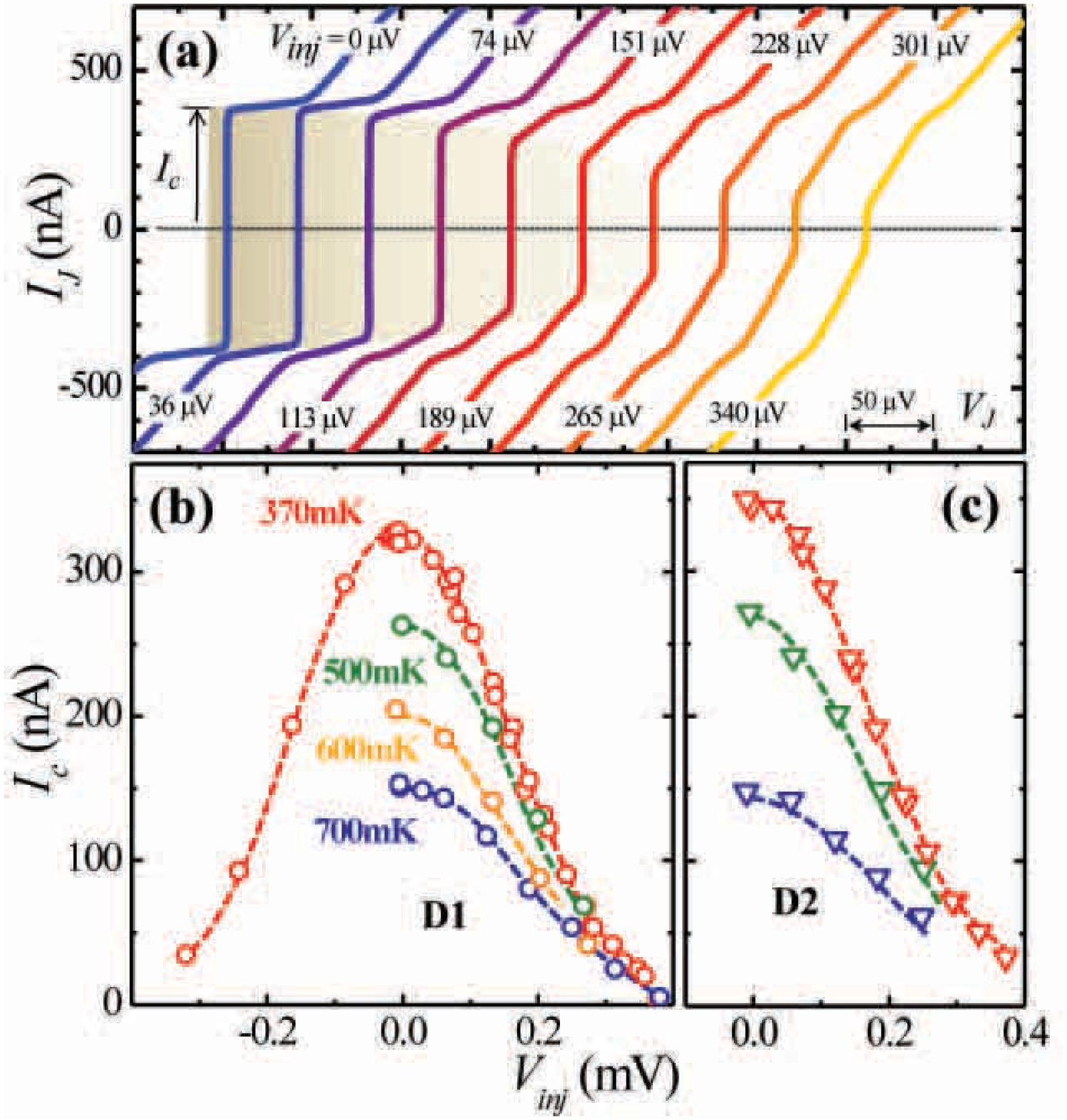}
\caption{{\bf Dependence of Josephson current on quasiparticle injection}.
(a) Injection of quasiparticles at the Ti/Au electrodes is very effective in tuning the supercurrent flow between the S leads.
Few hundred of $\,{\rm \mu V}$ are sufficient to completely quench the Josephson coupling, as it appears in these selected current-voltage characteristics of device D2 measured at $T=370\,{\rm mK}$. Hatched area provides a guide to the eye which emphasizes the supercurrent suppression as a function of $V_{inj}$.
The curves are horizontally shifted for clarity.
(b) and (c) Detailed evolution of the measured critical current $I_c$ as a function of the injection voltage $V_{inj}$ at different bath temperatures $T$ for devices D1 [(b)] and D2 [(c)]. Dashed lines are guides to the eye.
Datapoints shown in (b) for negative values of $V_{inj}$ indicate that Josephson current suppression is achieved in a symmetric fashion, regardless the sign of the injection current.}
\end{center}
\end{figure}
We emphasize that the observation and control of such a robust Josephson effect is made possible by the chosen device architecture, which allows us to use {\em short} SNS junctions while retaining full tunability of the supercurrent flow. In fact there is no fundamental lower limit to $L$ in our scheme, while other approaches such as, for instance, Josephson field effect transistors (JoFETs) \cite{Doh05,Clark80,Akazaki96}, require longer junctions to allow electrostatic gating and avoid excessive screening from the electrodes. These requirements lead to larger control voltages, weaker proximity effect and stray capacitance dictated by the integration of strongly-coupled gates on top of short junctions. Clearly the present operational principle is amenable to all-metal implementations. We should like to stress, however, that our NW-based architecture offers a significant beneficial feature, thanks to the resulting higher impedance in the injection line. The latter reduces the required control power by $\sim 2-3$ orders of magnitude with respect to what shown so far in comparable metallic systems~\cite{Morpurgo98,Baselmans99,Baselmans2001}.
\begin{figure}[t!]
\begin{center}
\includegraphics[width=0.5\textwidth]{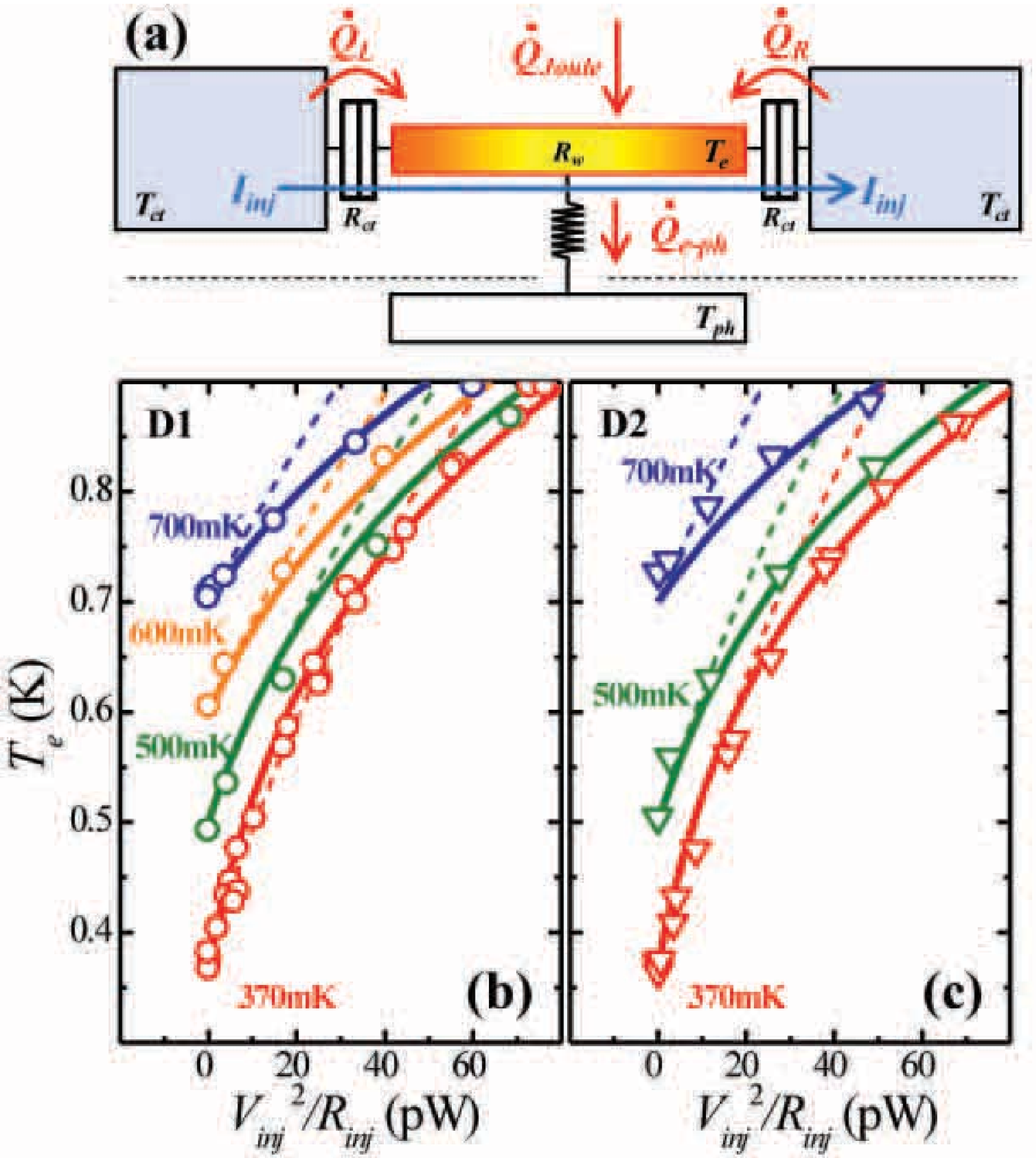}
\caption{{\bf Thermal budget analysis of the InAs NW Josephson transistors}.
(a) The temperature increase in the electron system of the InAs NW caused by quasiparticle injection is determined by the steady-state balance between in-coming and out-going heat flows in the device. The key ones are: Joule heating ($\dot{Q}_{Joule}$), heat leakage through the metallic contacts ($\dot{Q}_{L,R}$) and heat exchange between electrons and acoustic phonons ($\dot{Q}_{e-ph}$). The NW phonon subsystem is assumed to reside at the same temperature as the thermal bath. $T_{ph}$ and $T_{ct}$ represent the phonon and N leads temperatures, respectively, whereas $R_w$ and $R_{ct}$ denote the resistances of the NW and the contacts, respectively.
Panels (b) and (c): using the equilibrium $I_c$ vs T characteristics shown in Fig.~2 as a thermometer, we extract the electron temperature $T_e$ as a function of the injection bias $V_{inj}$ at different bath temperatures $T$ for device D1 [(b)] and D2 [(c)]. $R_{inj}$ is the total resistance of the injection line ($1.05\,{\rm k\Omega}$ and $1.28\,{\rm k\Omega}$ for device D1 and D2, respectively).
Based on statistical data acquired on several junctions of different length we estimated the contact resistance to be $R_{ct}\simeq100\,{\rm \Omega}$.
The experimental behavior is then compared to the phenomenological model sketched in panel (a).
The best fit is indicated by the continuous lines whereas the dashed ones correspond to the limit where electron-phonon interaction is neglected.
}
\end{center}
\end{figure}

Insight into the working principle of this device can be gained by recalling that in diffusive SNS junctions a continuum spectrum of Andreev bound states is responsible for the supercurrent flow across the structure \cite{Wilhelm,Heikkila}. Occupation of these supercurrent-carrying states is determined by the quasiparticle energy distribution in the N region which can, in turn, be controlled through the application of a finite bias $V_{inj}$ across the NW \cite{Wilhelm,Heikkila}. Although a detailed quantitative description of this effect is rather complex, we believe that the dominant physics governing our devices can be captured by considering the experimentally-observed heating of the electron system following current injection. Comparison between Figs.~2a and 3a suggests that upon hot-quasiparticle injection NW electrons can be described by a quasi-equilibrium Fermi-energy distribution characterised by a temperature $T_e$ larger than the bath temperature~\cite{Giazotto06}.
Thanks to the experimental data available, we can convert the measured $I_c(V_{inj})$ values into $T_e(V_{inj})$ by using the measured temperature-calibration data in Fig.~2b, i.e., by exploiting the equilibrium Josephson junction as an electron thermometer \cite{Savin04,Tirelli08,SNSThermo}.

The resulting evolution of the electron temperature $T_e$ is displayed in Fig.~4b and 4c for devices D1 and D2, respectively.
$T_e$ values as a function of $V_{inj}$ and $T$ are determined by the steady-state thermal balance of several in-coming and out-going heat fluxes within the device system, as sketched in Fig.~4a.
In particular, injection of hot carriers and thermal leakage through the lateral N electrodes provide a heat inflow contribution $\dot{Q}_L=\dot{Q}_R=V_{ct}^2/2R_{ct}-L(T_e^2-T_{ct}^2)/2R_c$, where the parameter $L$ is ideally equal to the Lorentz number $L_0=2.44\times10^{-8}\,{\rm W\Omega/K^2}$ and $V_{ct}$, $T_{ct}$ and $R_{ct}$ represent the voltage drop, the temperature and the contact resistance, respectively. Although current injection in the control line can in principle lead to a temperature gradient along the NW, we assume the latter to be described as lumped resistor $R_w$ characterized by an average electronic temperature $T_e$.
Furthermore, Joule dissipation within the NW itself is taken into account through the additional term $\dot{Q}_{Joule}=V_w^2/R_w$, where $V_w$ represents the voltage drop across the NW. These contributions can be recast in terms of the measured injection voltage $V_{inj}$ and total resistance of the injection line $R_{inj}$ by defining the lever arms $V_{ct}=\alpha_{ct}V_{inj}$ and $V_w=\alpha_wV_{inj}$ with the constraint $\alpha_w=1-2\alpha_{ct}$.
Similarly, it is possible to define $R_{ct}=\alpha_{ct}R_{inj}$ and $R_w=\alpha_wR_{inj}$. This leads to
\begin{equation}
\dot{Q}_L+\dot{Q}_R+\dot{Q}_{Joule}=
\frac{(1-\alpha_c)V_{inj}^2}{R_{inj}} - \frac{L(T_e^2-T^2)}{\alpha_cR_{inj}},
\end{equation}
where we set $T_{ct} = T$.
The other main relaxation mechanism which is expected to play a role in our structures is heat exchange between electrons and acoustic phonons ($\dot{Q}_{e-ph}$) in the NW. To this end we make the simplifying assumption that $\dot{Q}_{e-ph}= \Sigma V (T_e^5-T_{ph}^5)$, which holds for bulk metals at low temperatures (i.e., typically below $\sim 1$ K), where $\Sigma$ is a material-dependent constant~\cite{Giazotto06}, $V$ is the NW volume and $T_{ph}$ is the phonon temperature in the nanostructure.
Since 70-80\% of the NWs are covered by electrodes (see Fig.~1a), we also assume that $T_{ph} = T$, i.e., NW phonons thermalize at the Si-substrate and N-lead common bath temperature. Based on all this, we use
\begin{equation}
\dot{Q}_L+\dot{Q}_R+\dot{Q}_{Joule} - \dot{Q}_{e-ph} = 0
\label{TBE}
\end{equation}
to fit the data collected on devices D1 and D2.

Equation (\ref{TBE}) well describes the observed behaviour, as visible in Fig.~4b and 4c. $\alpha_{ct}$ is estimated $\simeq0.1$ based on transport data on wires of different lengths; we note that its value rescales the extracted best-fitting parameters $L$ and $\Sigma V$, but nevertheless does not modify the best-fit curves. We obtain $L=5.1\pm 0.5\times10^{-9}\,{\rm W\Omega/K^2}$ and $\Sigma=4.7\pm0.3\times10^9\,{\rm W/K^5m^3}$, using the NW volume $V=1.6\times 10^{-20}\, {\rm m^3}$. We note that our estimate of $\Sigma$ is comparable to values typical of bulk metallic films~\cite{Giazotto06}. Furthermore, although the $T^2$ term in Eq. 2 cannot properly account alone for the observed behaviour (see dashed lines in Figs. 4b and 4c), the precise relative contribution of lead and phonon thermal leakage is rather delicate to estimate owing to the correlation between the two fitting parameters.

We should finally like to comment on two important figures of merit of the transistor: its power gain and speed of operation. Even if the former is lacking in the present non-optimized devices, it can be made sizable by tuning the ratio between $R_{inj}$ and $R_N$ \cite{Wilhelm}, i.e. by fabricating sufficiently-long control lines. Concerning speed, switching time $\tau$ is determined by how fast an out-of-equilibrium electron energy distribution can be established in the NW \cite{Morpurgo98,Giazotto04}; in our case we estimate $\tau =L_{inj}^2/D\approx 50\,{\rm ps}$, which can be decreased by shortening $L_{inj}$ or increasing $D$.

In light of possible applications, the present InAs NW-based hot-electron Josephson transistor concept offers a number of attracting features: (i) a large, yet tunable, critical supercurrent; (ii) the possibility to achieve large power gains in optimized devices; (iii) a high operating frequency (easily up to $\sim 100$ GHz); (iv) a reduced dissipation ($\sim 100$ pW or below), ideal for low-temperature applications; (v) the possibility to combine the Josephson effect with custom quantum structures defined during the NW growth. The adoption of self-assembled nanostructures as the core element of this class of devices thus paves the way to rather broad opportunities since it allows the design and growth of epitaxial quantum systems characterized by atomically-defined potential profiles~\cite{Roddaro08,Jiang07}, a crucial element for several innovative out-of-equilibrium nanodevice schemes~\cite{Giazotto06,Tirelli08,Linke09}.

We gratefully acknowledge R. Fazio, H. Linke, J. Matthews, F. Taddei and H. Xu for fruitful discussions. The work for partially supported by the INFM-CNR Seed projects ``Acoustoelectrics on Self-Assembled One-dimensional Semiconductors'' and ``Quantum-Dot Refrigeration: Accessing the $\mu$K Regime in Solid-State Nanosystems'', and by the NanoSciERA project ``NanoFridge''.

\section{METHODS}

{\bf Fabrication details of InAs NW devices and measurement set-up.}
Selenium doped NWs were grown by chemical beam epitaxy on a InAs 111B substrate. Gold catalyst particles were formed by thermal dewetting (at 520$^\circ$C for 20 min) of a 0.5 nm thick Au film under TBA flux. NWs were grown for 2 hours at $420\,{\rm ^\circ C}$ using TBA ($2.0\,{\rm Torr}$), TMI ($0.3\,{\rm Torr}$) and DTBSe ($0.4\,{\rm Torr}$). NWs had typically a diameter $d=90\pm10\,{\rm nm}$ and length $\simeq 2.5\,{\rm \mu m}$. Transport parameters were estimated over an ensemble of nominally identical $1\,{\rm \mu}$m-long NW field effect transistors using a charge control model~\cite{Jiang07} and a numerical evaluation of the gate capacitance. Carrier density was estimated to be $n = 1.8\pm0.8\times 10^{19}\,{\rm cm^{-3}}$ and electron mobility $\mu = 300\pm 100\,{\rm cm^2/Vs}$. Hot-electron Josephson transistors were fabricated using a technique of dry cleavage of the NWs onto Si/SiO$_2$ substrates ($500\,{\rm nm}$ oxide on intrinsic Si). Contacts were obtained by a two-step aligned process: thermal evaporation of Ti/Au ($10/80\,{\rm nm}$) was performed first and followed by e-beam deposition of Ti/Al ($10/80\,{\rm nm}$).
In order to obtain transparent metal-NW contacts great care was devoted to the passivation of the semiconductor surface : InAs NWs were treated with a NH$_4$S$_x$ solution before each evaporation step. The electric characterization was performed by cooling the devices in a filtered $^3$He refrigerator down to about $370\,{\rm mK}$. Current injection at the normal contacts was obtained using a floating biasing source. Currents and voltages were measured using room-temperature preamplifiers.

{\bf Theoretical model for a diffusive SNS Josephson junction in the short limit.}
The measured Josephson critical currents were analyzed by using a model which holds for short diffusive SNS junctions. In this regime the critical supercurrent $I_c$ is expected to scale with the temperature as~\cite{Heikkila,ShortDiffusive}
\begin{eqnarray}
I_c(T,\phi) &=&\beta\frac{2\pi\Delta(T) k_BT\cos(\phi/2)}{qR_N} \\ \nonumber
&\times& \sum_n\frac{1}{\Omega_n(T,\phi)}\tan^{-1}\frac{\Delta(T)\sin(\phi/2)}{\Omega_n(T,\phi)}
\end{eqnarray}
where $\Omega_n(T,\phi)=\sqrt{\Delta^2(T)\cos^2(\phi/2)+\omega_n^2(T)}$, $\omega_n(T)=\pi k_BT(2n+1)$, and $\beta$ is a phenomenological suppression coefficient which accounts for the non-ideality of the junction~\cite{Courtois08,Garcia09}. In the above equation, $k_B$ is the Boltzmann constant, $q$ is the electron charge, $R_N$ is the normal-state resistance of the junction, $\Delta$ is the superconducting order parameter, and $\phi$ is the macroscopic phase difference across the SN boundaries.

\section{Supplementary materials}

\begin{figure}[t!]
\begin{center}
\includegraphics[width=0.48\textwidth]{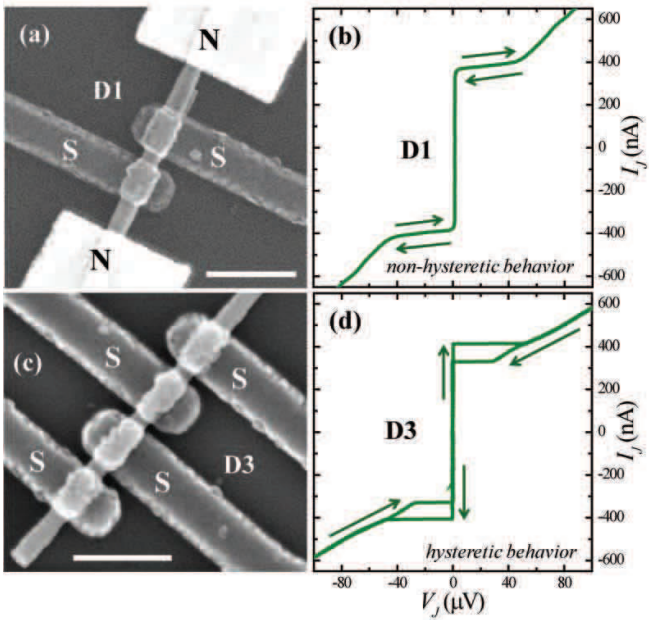}
\caption{{\bf Supplementary Fig. S1. Thermal sinks in the Josephson NW transistors}.
Non-hysteretic (b) or hysteretic (d) current-voltage characteristics can be obtained depending on the type of contact electrode fabricated on top of the device. The different phenomenology is consequence of the marked difference in terms of thermal sinking of N contacts (a) with respect to S ones (c). The characteristics shown in (b) and (d) were measured at $T=370$ mK on the device shown in (a) and (c), respectively.}
\end{center}
\end{figure}
Metallic contacts play a crucial role in thermal sinking of the NW device \cite{Giazotto06}. Figure~S1 shows the different behaviour observed on Josephson NW transistors containing either both N and S contacts (Fig. S1 a), or only S ones (Fig. S1 c). In the first case (see Fig. S1 b) we never observed hysteresis in the IV characteristics, while in the second case (see Fig. S1 d) a clear hysteresis developed for devices which exhibit $I_c$ exceeding $\sim 100\,{\rm nA}$.
This different behaviour, which was verified on a number of devices, can be ascribed to the poor thermal conductivity of the superconducting leads in good metallic contact with the NW which act as Andreev {\em mirrors} for what concerns electronic heat conduction \cite{Giazotto06}.
Hysteresis in mesoscopic SNS junctions is believed to stem from electron heating of the N region once the junction switches to the resistive branch \cite{Courtois08}. Our data suggest that the device in Fig. S1c heats up more than that in Fig. S1a when it switches to the dissipative regime.

The differential resistance spectra versus voltage of a typical Josephson NW transistor with all-superconducting leads is visible in Fig. S2a. The plot shows multiple Andreev reflection features \cite{Octavio83} occurring at values $V_J=2e\Delta/n$ with $n=1$, $2$, and $3$ for a sequence of different bath temperatures.
The extracted value of the gap parameter is $\Delta\simeq 120\,{\rm \mu eV}$. The $V_J$ position of the minima in $dV_J/dI_J$ as a function of the temperature (see Fig. S2b) closely follows the Bardeen-Cooper-Schrieffer prediction for the superconducting gap \cite{Tinkham}.
\begin{figure}[t!]
\begin{center}
\includegraphics[width=0.45\textwidth]{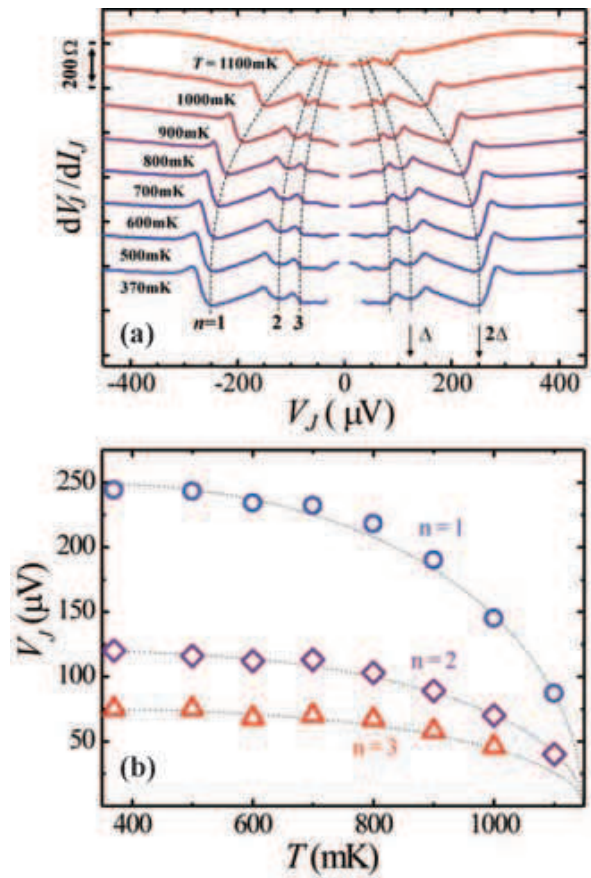}
\caption{{\bf Supplementary Fig. S2. Multiple Andreev reflections in the Josephson NW transistor}.
(a) Multiple Andreev reflections manifesting in the differential resistance vs voltage characteristics of an InAs S-NW-S junction at selected bath temperatures. Three to four reflection orders are typically visible in the studied devices. The extracted value for the superconducting gap is $\Delta\simeq 120\,{\rm \mu eV}$. (b) Position of the differential resistance minima for $n=1$, $2$ and $3$ as a function of the temperature, and comparison with the Bardeen-Cooper-Schrieffer prediction (dashed lines) for $\Delta(T)$.}
\end{center}
\end{figure}

\end{document}